\documentclass[prc,nofootinbib,superscriptaddress,twocolumn,showpacs,amsmath,amssymb]{revtex4}

\usepackage{graphicx}

\renewcommand{\vec}[1]{\mbox{\boldmath $#1$}}

\begin{document}

\title{Dipole excitation and geometry of borromean nuclei}

\author{K. Hagino}
\affiliation{ 
Department of Physics, Tohoku University, Sendai, 980-8578,  Japan} 

\author{H. Sagawa}
\affiliation{
Center for Mathematical Sciences,  University of Aizu, 
Aizu-Wakamatsu, Fukushima 965-8560,  Japan}


\begin{abstract}
We analyze the Coulomb breakup cross sections of $^{11}$Li and $^6$He 
nuclei 
using a three-body model with a density-dependent contact
interaction. We show that the concentration of the B(E1) strength 
near the threshold can be well reproduced with this model. 
With the help of the calculated B(E1) value, we extract the
root-mean-square (rms) distance between the core nucleus and
the center of mass of two valence neutrons without resorting to 
the sum rule, which may suffer from unphysical Pauli forbidden
transitions. Together with  the empirical rms distance 
between the neutrons obtained 
from the matter radius study and also from 
the three-body correlation study in the break-up reaction, 
we convert these rms distances to the mean opening 
angle between the valence neutrons from the core nucleus. 
We find that the obtained 
mean opening angles in  $^{11}$Li and $^6$He 
agree with the
three-body model predictions.
\end{abstract}


\maketitle

It has been well recognized by now that 
weakly bound nuclei exhibit a strong multipole strength which is 
concentrated near the continuum threshold, because of the optimal 
matching of wave functions between a weakly bound and continuum 
states \cite{BW52,BBH91,STG92,NLV05,TB05}. 
Recently, Nakamura {\it et al.} have remeasured the low-lying 
dipole excitations in $^{11}$Li nucleus and have confirmed for the 
first time the strong concentration of the dipole 
strength near the threshold in this 2-neutron (2n) halo nucleus
\cite{N06}. The low-lying dipole strength for another
2n halo nucleus, $^6$He, has also been measured by 
Aumann {\it et al.} \cite{A99}. 

As well as being of considerable interest in its own right, 
the B(E1) strength distribution of 2-neutron halo nuclei is 
also important as it is intimately related to the root-mean-square
(rms) distance, $\sqrt{\langle r^2_{c-2n}\rangle}$, 
between the core nucleus and the center-of-mass of two
valence neutrons \cite{BF1,BF2,EHMS07}. 
Together with an additional information for the rms distance between 
the two neutrons, 
$\sqrt{\langle r^2_{nn}\rangle}$, 
one can then extract the geometry of 2n halo 
nuclei, such as the mean opening angle between the neutrons from the core
 \cite{N06,BH07}. 
This information is particularly important to extract the strength of
di-neutron correlations in halo nuclei.

In the previous applications, 
the rms distance $\sqrt{\langle r^2_{c-2n}\rangle}$ has
been obtained from the measured B(E1) strength using the 
relation \cite{BF1,BF2,EHMS07}, 
\begin{equation}
B(E1)=\frac{3}{\pi}\left(\frac{Ze}{A}\right)^2\,
\langle r^2_{c-2n}\rangle. 
\end{equation}
This relation is obtained with closure, which includes unphysical
Pauli forbidden transitions to the states with negative excitation
energies. 
Although the effect of Pauli forbidden
transitions is not large, it leads to a non-negligible correction. 
In Ref. \cite{EHMS07}, a better prescription has been
proposed recently, which uses a model calculation for the B(E1) value 
and $\langle r^2_{c-2n}\rangle$ to extract the ``experimental value'' 
for $\langle r^2_{c-2n}\rangle$ (See Eq. (6) in Ref. \cite{EHMS07},
and Eq. (\ref{r-c2n}) below). 
Although this prescription uses theoretical values, it has been shown
that the model dependence is insignificant \cite{EHMS07}. 

The aim of this paper is to analyze the mean opening angle of
valence neutrons in the 2n halo nuclei, $^{11}$Li and $^6$He, using several
empirical information. To this end, we first discuss 
the new prescription for $\langle r^2_{c-2n}\rangle$  
by analyzing  the Coulomb dissociation cross
sections of these nuclei with a three-body model. 
Assuming the three-body character, one can also extract the distance between two neutrons, 
 $\sqrt{\langle r^2_{nn}\rangle}$, 
 from  the empirical information of matter radii and
 $\langle r^2_{c-2n}\rangle$ 
(see Eq. (\ref{eq:rnn})). An alternative way to 
 extract the value  for $\sqrt{\langle r^2_{nn}\rangle}$ is the three-body 
correlation study in the dissociation of two neutrons in halo nuclei 
\cite{M00}.  We will discuss the two ways to determine the mean opening
angle by using these empirical information. 

The three-body model which we employ in this paper is exactly the same as in
Refs. \cite{HS05,HSCS07,BF3}.
The model adapts a 
density-dependent contact interaction between the valence 
neutrons \cite{BF1,BF2}. The recoil kinetic energy of the core 
nucleus is taken into account as in Ref. \cite{BF3}. 
Single particle continuum states are discretized by putting a nucleus
in a large box. 
The wave functions for the ground state with $J^{\pi}=0^+$ and 
for the excited $J^{\pi}=1^-$ states are then obtained by diagonalizing 
the three-body Hamiltonian within a large model space which 
is consistent with the $nn$ interaction. 
We use the same values for the parameters as in Ref. \cite{HS05}. 

\begin{figure}
\includegraphics[scale=0.5,clip]{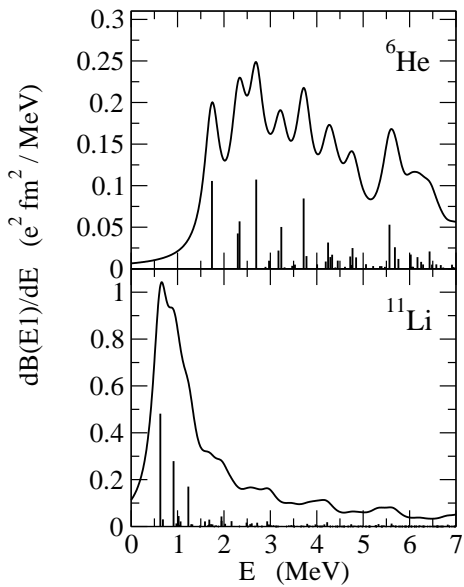}\\
\caption{
The B(E1) distribution for the $^6$He and $^{11}$Li nuclei. 
The solid curve is obtained with a smearing procedure with 
$\Gamma$=0.2 MeV. } 
\end{figure}

The dipole strength distributions for the $^6$He and $^{11}$Li nuclei 
obtained with this model are shown in Fig. 1
\footnote{
We have found that 
the matrix elements for the off-diagonal part of the recoil kinetic energy 
were not properly evaluated for the $J^{\pi}=1^-$ states 
in Fig. 5 of Ref. \cite{HS05}. 
This technical problem 
 has been 
cured 
in Fig. 1 of  the present manuscript (it had been cured 
in  the B(E2) calculations of 
$^{16}$C in Ref. \cite{HS07}). 
 Although this error did not cause any 
substantial change in  $^{11}$Li, 
the B(E1) distribution for $^6$He is considerably different from
the previous calculation. 
The energy of the first peak in
the B(E1) distribution is now at 1.75 and 0.66 MeV 
for $^6$He and $^{11}$Li, respectively 
(previously, it was at 1.55 and 0.66 MeV 
for $^6$He and $^{11}$Li, respectively). 
These values are still close to 1.6\,$S_{2n}$, where
$S_{2n}$ is the 2n separation energy, and our conclusion that the 
dineutron correlation plays an important role in these nuclei remains
unchanged. 
}. 
Also shown by the solid
curves are the B(E1)
distributions smeared with the Lorenzian function with the width of
$\Gamma=0.2$ MeV. 
For the $^6$He nucleus, we obtain the total B(E1) strength of 0.660
e$^2$fm$^2$ up to $E\leq 5$ MeV and 1.053 
e$^2$fm$^2$ up to $E\leq 10$ MeV. 
These are in good agreement with the 
experimental values, 
B(E1; $E\leq$ 5 MeV)=0.59 $\pm$ 0.12 e$^2$fm$^2$ and 
B(E1; $E\leq$ 10 MeV)=1.2 $\pm$ 0.2 e$^2$fm$^2$ \cite{A99}. 
For the $^{11}$Li nucleus, we obtain the total B(E1) strength of 1.405 
e$^2$fm$^2$ up to $E_{\rm rel}=E-S_{2n} \leq 3$ MeV, which is 
compared to the experimental value, 
B(E1; $E_{\rm rel}\leq$ 3 MeV)=1.42 $\pm$ 0.18 e$^2$fm$^2$
\cite{N06}. Again, the experimental data is well reproduced within the 
present model. 
In Ref. \cite{EHMS07}, it has been proposed to estimate the
experimental value for $\langle r^2_{c-2n}\rangle$ using the relation,
\begin{equation}
\langle r^2_{c-2n}\rangle_{\rm exp} 
= \frac{B(E1; E\leq E_{\rm max})_{\rm exp}}
{B(E1; E\leq E_{\rm max})_{\rm cal}}\,
\cdot\langle r^2_{c-2n}\rangle_{\rm cal}. 
\label{r-c2n}
\end{equation}
From the calculated values for 
$\langle r^2_{c-2n}\rangle_{\rm cal}$, that is, 13.2 and 26.3 fm$^2$
for $^6$He and $^{11}$Li, respectively, we thus obtain 
$\sqrt{\langle r^2_{c-2n}\rangle_{\rm exp}}=3.878 \pm 0.324$ fm and 
5.15 $\pm$ 0.327 fm for $^6$He and $^{11}$Li, respectively. 
Notice that the value for the $^6$He nucleus is somewhat larger than
the one estimated in Ref. \cite{A99}, that is, 3.36 $\pm$ 0.39 fm. 

\begin{figure}
\includegraphics[scale=0.5,clip]{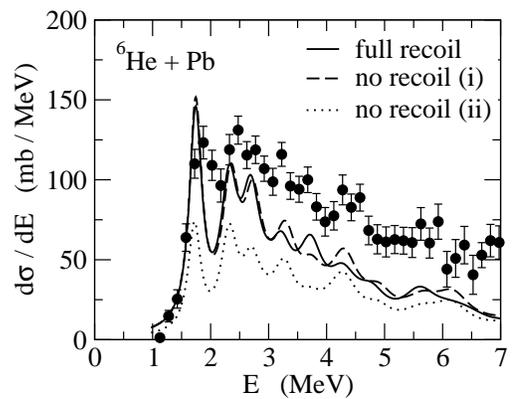}\\
\caption{
Coulomb breakup cross sections for $^6$He+Pb at 240 MeV /nucleon. 
The solid line is the result of the full three-body calculations,
while the dashed line is obtained by neglecting the off-diagonal
component of the recoil kinetic energy in the excited states. 
The dotted line is obtained by neglecting the off-diagonal recoil term
both in the ground and the excited states. 
These results are smeared with an energy dependent width of 
$\Gamma = 0.15 \cdot \sqrt{E_{\rm rel}}$ MeV. 
The experimental data are taken from Ref. \cite{A99}. 
}
\end{figure}

\begin{figure}
\includegraphics[scale=0.5,clip]{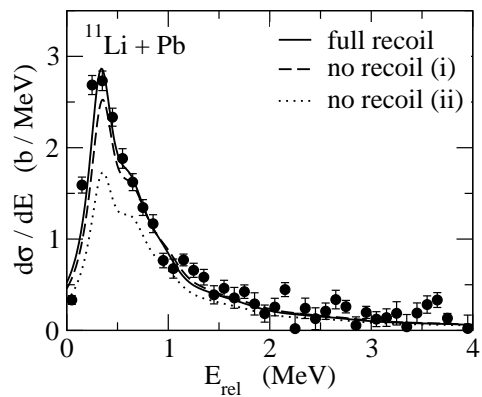}\\
\caption{
Coulomb breakup cross sections for the $^{11}$Li +Pb at 70 MeV/nucleon.
The meaning of each line is the same as in Fig. 2. 
The calculated results are smeared with an energy dependent width of 
$\Gamma = 0.25 \cdot \sqrt{E_{\rm rel}}$ MeV. 
The experimental data are taken from Ref. \cite{N06}. 
}
\end{figure}

We next evaluate  the Coulomb breakup cross sections, especially
 paying attention to the recoil effect of the core nucleus.  
Based on the relativistic Coulomb excitation theory 
\cite{BB88,WA79},  the cross sections  are
obtained by multiplying the virtual photon number $N_{\rm E1}(E)$ 
to the B(E1) distribution shown in Fig. 1. The solid curves in Figs. 2
and 3 show  the Coulomb breakup cross sections thus obtained for
$^6$He+Pb reaction at 240 MeV/nucleon \cite{A99} and 
$^{11}$Li+Pb reaction at 70 MeV/nucleon \cite{N06}, respectively. 
In order to facilitate the comparison with the experimental data, we 
smear the discretized cross sections with the Lorenzian function 
with an energy dependent width, $\Gamma = \alpha\cdot \sqrt{E_{\rm rel}}$. 
We take $\alpha=0.15$ and 0.25 MeV$^{1/2}$ for $^6$He and 
$^{11}$Li, respectively. We see that the experimental breakup cross
sections are reproduced remarkably well within the present three-body
model, especially for the $^{11}$Li nucleus. 

One of the advantages to use the contact interaction is that the
continuum response can be calculated relatively easily \cite{BF2}. 
In the presence of the recoil kinetic energy of the core nucleus,
however, this advantage disappears since the off-diagonal part of
the recoil energy, $\vec{p}_1\cdot\vec{p}_2/(A_cm)$ (the last term 
in Eq. (1) of Ref. \cite{HS05}) is a finite range two-body term,
although the diagonal part, $(\vec{p}^2_1+\vec{p}^2_2)/(2A_cm)$, 
can be easily included through the reduced mass. 
In order to examine  
the effect of the recoil term, 
Figs. 2 and 3 compare 
the Coulomb breakup cross sections calculated 
 by taking into account  
the recoil term exactly (the solid curves) with those calculated approximately (the dashed 
and dotted curves). 
For the dashed curves, the off-diagonal component of the recoil kinetic
energy is neglected in the excited $J^\pi=1^-$ states, while it is fully taken 
into account in the ground state. 
It is interesting to notice that these calculations lead to similar 
results to the one in which the recoil term is treated exactly (the
solid curves). The continuum response was obtained in this way in
Ref. \cite{EHMS07} for the $^{11}$Li nucleus. 
The dotted curves, on the other hand, are obtained by neglecting the 
off-diagonal part of the recoil term both for the ground and the 
$J^\pi=1^-$ states. For this calculation, we slightly readjust the
parameters of the pairing interaction so that the ground state energy 
remains the same. By neglecting the recoil term in the ground state, 
the value for $\langle r^2_{c-2n}\rangle$ decreases, 
from 13.2 fm$^2$ to 9.46 fm$^2$ for $^6$He 
and 
from 26.3 fm$^2$ to 20.58 fm$^2$ for $^{11}$Li. Consequently, 
the B(E1) distribution as well as the breakup cross sections are
largely underestimated. The fraction of the main components in the ground
state wave function are also altered by neglecting the recoil term: 
for $^6$He, the fraction of the (p$_{3/2})^2$ component changes from 
83.0 \% to 90.8 \%, 
and for $^{11}$Li, the fraction of the (s$_{1/2})^2$ component changes from 
22.6\% to 17.1\%, 
and the fraction of the (p$_{1/2})^2$ component from 
59.1 \% to 65.7 \%. These results clearly indicate that the 
recoil term is important for the ground state, 
 while it has a rather small effect on the excited states.

\begin{figure}
\includegraphics[scale=0.5,clip]{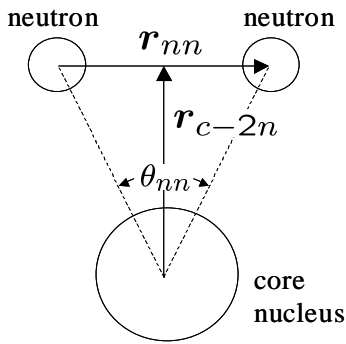}\\
\caption{
Geometry of a 2n halo nucleus consisting of a core nucleus and 
two valence neutrons. 
}
\end{figure}

Let us now discuss the geometry of the $^6$He and $^{11}$Li nuclei. 
Using the experimental value for $\langle r^2_{c-2n}\rangle$ obtained 
from the B(E1) distribution, one can extract the mean opening angle 
between the valence neutrons once an additional information is
available. 
In Ref. \cite{N06}, Nakamura {\it et al.} have used the non-correlated 
calculation for B(E1) distribution given in Ref. \cite{BF2} to
estimate the rms distance between the core nucleus and one of the
valence neutrons, 
and obtained the mean opening angle of $\langle\theta_{nn}\rangle 
= 48^{+14}_{-18}$ 
degrees. However, this method is highly model dependent and also it is not 
obvious whether the non-correlated calculation is reasonable to
estimate the rms distance. 
The mean opening angle can be extracted more directly when the rms 
distance between the valence neutrons, 
$\langle r^2_{nn}\rangle$, is available (see Fig. 4). This quantity is related 
to the matter radius and $\langle r_{c-2n}^2\rangle$ 
in the  three-body model \cite{BF1,BF3,EHMS07,VMP96}, 
\begin{equation}
\langle r_m^2\rangle = 
\frac{A_c}{A}\,\langle r_m^2\rangle_{A_c}
+\frac{2A_c}{A^2}\,\langle r_{c-2n}^2\rangle 
+\frac{1}{2A}\,\langle r_{nn}^2\rangle, 
\label{eq:rnn}
\end{equation}
where $A_c=A-2$ is the mass number of the core nucleus. 
The matter radii 
$\langle r_m^2\rangle$ can be estimated from interaction cross
sections. Employing the Glauber theory in the optical limit, Tanihata
{\it et al.} have obtained 
$\sqrt{\langle r_m^2\rangle}$ = 1.57 $\pm$ 0.04, 
2.48 $\pm$ 0.03, 2.32 $\pm$ 0.02, and 
3.12 $\pm$ 0.16 fm for $^4$He, $^6$He, $^9$Li, and 
$^{11}$Li, respectively \cite{T88}. Using these values, we obtain the 
rms neutron-neutron distance of 
$\sqrt{\langle r_{nn}^2\rangle}$ = 3.75 $\pm$ 0.93 and 
5.50 $\pm$ 2.24 fm for $^6$He and $^{11}$Li, respectively. 
Combining these values with the rms core-dineutron distance, 
$\sqrt{\langle r_{c-2n}^2\rangle}$, obtained with Eq. (\ref{r-c2n}),
we obtain the mean opening angle of $\langle\theta_{nn}\rangle$ =
51.56$^{+11.2}_{-12.4}$ and 56.2$^{+17.8}_{-21.3}$ degrees for 
$^6$He and $^{11}$Li, respectively. 
These values are comparable to 
the result of the three-body model calculation, 
$\langle\theta_{nn}\rangle$=66.33 and 65.29 degree for $^6$He and 
$^{11}$Li, respectively \cite{HS05}, although 
the experimental values are somewhat smaller.
We should remark here   
that it is misleading to say that 
the two neutrons are mostly sitting with an opening angle obtained 
in this way. Instead, the mean opening angle is most probably 
an average of a smaller and
a larger correlation angles in the density distribution as has been suggested 
in Ref. \cite{HS05}. 

\begin{table}[hbt]
\caption{
The geometry of the $^6$He and $^{11}$Li nuclei 
extracted from various experimental data. 
The mean opening angles calculated with the three-body model \cite{HS05} are given in 
the last line for each nucleus in the table. 
}
\begin{center}
\begin{tabular}{c|c|cc|c}
\hline
\hline
nucleus & $\sqrt{\langle r_{c-2n}^2\rangle}$ & 
$\sqrt{\langle r_{nn}^2\rangle}$ & method & 
$\langle\theta_{nn}\rangle$ \\
& (fm) & (fm) & & 
  (degree) \\
\hline
$^6$He & 3.88$\pm$0.32 & 
3.75 $\pm$ 0.93 & (matter radii) & 51.6 $^{+11.2}_{-12.4}$ \\
&  & 
5.9$\pm$ 1.2 & (2n-correlations) & 74.5 $^{+11.2}_{-13.1}$ \\
   & &      &  &66.33 \cite{HS05}\\
\hline
$^{11}$Li & 5.15$\pm$0.33 & 
5.50 $\pm$ 2.24 & (matter radii) & 56.2 $^{+17.8}_{-21.3}$ \\
&  & 
6.6$\pm$ 1.5 & (2n-correlations) & 65.2 $^{+11.4}_{-13.0}$ \\
   & &      &   & 65.29 \cite{HS05} \\\hline
\hline
\end{tabular}
\end{center}
\end{table}

An alternative way to 
 extract the value  for $\sqrt{\langle r^2_{nn}\rangle}$ has been
 proposed which uses 
the three-body 
correlation study in the dissociation of two neutrons in halo nuclei 
\cite{M00}.  The two neutron correlation function provides the experimental
values for $\sqrt{\langle r_{nn}^2\rangle}$ to be  5.9 $\pm$ 1.2 and 6.6 $\pm$ 1.5
fm for $^6$He, $^{11}$Li, respectively \cite{M00}.
Very recently, Bertulani and Hussein used these values
to estimate the mean opening
angles and 
obtained 
$\langle\theta_{nn}\rangle$=83 $^{+20}_{-10}$ and 
66 $^{+22}_{-18}$ degrees 
for $^6$He and $^{11}$Li, respectively \cite{BH07}. 
When one adopts the presently obtained value
 for $\sqrt{\langle r_{c-2n}^2\rangle}$ 
with Eq. (\ref{r-c2n}) instead of those in Refs. \cite{N06,A99}, 
one  obtains  
$\langle\theta_{nn}\rangle$=74.5 $^{+11.2}_{-13.1}$ and 
65.2 $^{+11.4}_{-13.0}$ degrees 
for $^6$He and $^{11}$Li, respectively. Notice that these values are
in better agreement with the results of the three-body calculation
\cite{HS05}, especially for the $^6$He nucleus, as compared to the
values obtained by Bertulani and Hussein. 
We summarize our results in Table I.  One should notice that there are 
still large uncertainties in the empirical values of
$\sqrt{\langle r_{nn}^2\rangle}$ 
and,  consequently,  in the average opening angles 
$\langle\theta_{nn}\rangle$ as listed in Table 1.
 It is still an open challenging problem to determine experimentally the 
 values for $\sqrt{\langle r_{nn}^2\rangle}$  with 
higher precision. 

In summary, we have used the three-body model with a density dependent 
contact interaction to analyze the B(E1) distributions as well as the
Coulomb breakup cross sections of the $^6$He and $^{11}$Li nuclei. 
We have shown that the strong concentration of the B(E1) strength near 
the continuum threshold can be well reproduced with the present model
for both the nuclei. We have also 
examined the recoil effect 
of the core nucleus on the Coulomb breakup cross sections.
It is shown that
the recoil effect plays an important role
in the ground state while it may
be neglected in the excited states.  
Using the calculated B(E1) strength, we extracted 
the experimental value for the rms distance between the 
core and the center of two neutrons, 
which was then converted to the mean opening angle
of the two valence neutrons from the core nucleus. 
We have found that the mean opening angles thus obtained are in good
agreement with  the results of the three-body model calculation. 

\medskip

We thank T. Aumann and T. Nakamura for useful discussions on 
the experimental data of B(E1) strength distributions in 
the $^{6}$He and $^{11}$Li nuclei.
This work was supported by the Japanese
Ministry of Education, Culture, Sports, Science and Technology
by Grant-in-Aid for Scientific Research under
the program number 19740115.


\begin{thebibliography}{99}

\bibitem{BW52}J.M. Blatt and V.F. Weisskopf, {\it Theoretical 
Nuclear Physics}, (John Wiley \& Sons, New York, 1952). 

\bibitem{BBH91}C.A. Bertulani, G. Baur, and M.S. Hussein, 
Nucl. Phys. {\bf A526}, 751 (1991).

\bibitem{STG92}H. Sagawa, N. Takigawa, and Nguyen Van Giai, 
Nucl. Phys. {\bf A543}, 575 (1992).

\bibitem{NLV05}M.A. Nagarajan, S.M. Lenzi, and A. Vitturi, 
Eur. Phys. J. {\bf A24}, 63 (2005). 

\bibitem{TB05}S. Typel and G. Baur, Nucl. Phys. {\bf A759}, 247
  (2005). 

\bibitem{N06}T. Nakamura {\it et al.}, Phys. Rev. Lett. 
{\bf 96}, 252502 (2006). 

\bibitem{A99}T. Aumann {\it et al.}, Phys. Rev. C{\bf 59}, 1252 (1999). 

\bibitem{BF1}
G.F.  Bertsch and  H. Esbensen, Ann. Phys. (N.Y.) {\bf 209}, 327
(1991).

\bibitem{BF2}  H. Esbensen  and  G.F. Bertsch, Nucl. Phys. {\bf A542}, 310
  (1992).

\bibitem{EHMS07}H. Esbensen, K. Hagino, P. Mueller, and 
H. Sagawa, Phys. Rev. C{\bf 76}, 024302 (2007). 

\bibitem{BH07}C.A. Bertulani and M.S. Hussein, 
arXiv:0705.3998. 

\bibitem{M00}F.M. Marques {\it et al.}, Phys. Lett. B{\bf 476}, 219
  (2000). 

\bibitem{HS05}K. Hagino and H. Sagawa, Phys. Rev. C{\bf 72}, 
044321 (2005). 

\bibitem{HSCS07}K. Hagino, H. Sagawa, J. Carbonell, and P. Schuck, 
Phys. Rev. Lett. {\bf 99}, 022506 (2007). 

\bibitem{BF3} H. Esbensen, G.F.  Bertsch and K. Hencken,
  Phys. Rev. C{\bf 56}, 3054 (1999).

\bibitem{HS07}K. Hagino and H. Sagawa, 
Phys. Rev. C{\bf 75}, 021301 (R), (2007). 

\bibitem{BB88}C.A. Bertulani and G. Baur, Phys. Rep. {\bf 163}, 299
  (1988); Nucl. Phys. {\bf A480}, 615 (1988). 

\bibitem{WA79}A. Winther and K. Alder, Nucl. Phys. {\bf A319}, 518
  (1979). 

\bibitem{VMP96}N. Vinh Mau and J.C. Pacheco, Nucl. Phys. {\bf A607}, 
163 (1996). 

\bibitem{T88}I. Tanihata {\it et al.}, Phys. Lett. B{\bf 206}, 592
  (1988); A. Ozawa {\it et al.}, Nucl. Phys. {\bf A693}, 32 (2001). 




\end{thebibliography}
\end{document}